\documentclass[twocolumn,showpacs,preprintnumbers,amsmath,amssymb,prb]{revtex4}

\usepackage{graphicx}
\usepackage{dcolumn}
\usepackage{bm}

\begin{document}

\preprint{}

\title{
Manifestations of the hyperfine interaction between electron and
nuclear spins in singly-charged (In,Ga)As/GaAs quantum dots}

\author{Roman~V.~Cherbunin}
\affiliation{Experimentelle Physik II, Technische Universit\"{a}t
Dortmund, 44221 Dortmund, Germany} \affiliation{Institute of
Physics, St. Petersburg State University, St. Petersburg, 198504,
Russia}

\author{Thomas~Auer}
\affiliation{Experimentelle Physik II, Technische Universit\"{a}t
Dortmund, 44221 Dortmund, Germany}

\author{Alex~Greilich}
\affiliation{Experimentelle Physik II, Technische Universit\"{a}t
Dortmund, 44221 Dortmund, Germany}

\author{Ivan~V.~Ignatiev}
\email{ivan_ignatiev@mail.ru} \affiliation{Experimentelle Physik
II, Technische Universit\"{a}t Dortmund, 44221 Dortmund, Germany}
\affiliation{Institute of Physics, St. Petersburg State
University, St. Petersburg, 198504, Russia}

\author{Ruth~Oulton$^{\dag}$}
\affiliation{Experimentelle Physik II, Technische Universit\"{a}t
Dortmund, 44221 Dortmund, Germany}

\author{Manfred~Bayer}
\affiliation{Experimentelle Physik II, Technische Universit\"{a}t
Dortmund, 44221 Dortmund, Germany}

\author{Dmitri~R.~Yakovlev} \affiliation{Experimentelle Physik II,
Technische Universit\"{a}t Dortmund, 44221 Dortmund, Germany}
\affiliation{Ioffe Physico-Technical Institute, 194021 St.
Petersburg, Russia}

\author{Gleb G. Kozlov}
\affiliation{Institute of Physics, St. Petersburg State
University, St. Petersburg, 198504, Russia}

\author{Dirk~Reuter}
\affiliation{Angewandte Festk\"{o}rperphysik, Ruhr-Universit\"{a}t
Bochum, D-44780 Bochum, Germany}

\author{Andreas~D.~Wieck}
\affiliation{Angewandte Festk\"{o}rperphysik, Ruhr-Universit\"{a}t
Bochum, D-44780 Bochum, Germany}

\date{\today}

\begin{abstract}
The nuclear spin fluctuations (NSF) as well as the dynamic nuclear
polarization (DNP) and their effects on the electron spins in
negatively charged (In,Ga)As/GaAs quantum dots have been studied
by polarized pump-probe and photoluminescence spectroscopy
techniques. The effective magnetic field of the NSF is about 30~mT
at low excitation power. The NSF distribution becomes highly
anisotropic at strong optical excitation by circularly polarized
light with periodically alternating helicity. This phenomenon is
attributed to a decrease of the nuclear spin entropy due to the
hyperfine interaction with polarized electron spins. The DNP is
limited to small values for intense, but short photoexcitation.
\end{abstract}


\maketitle

\section{Introduction}

The strong localization of the electron wave function in a quantum
dot (QD) leads to a considerable increase of the electron density
at the nuclear sites and, therefore, enhances the hyperfine
interaction of the electron spin with the nuclear spins
\cite{BrownPRB96,GammonScience97,GammonPRL01}. The interaction
changes the electron and nuclear spin dynamics, and can be
described by a Fermi contact-type Hamiltonian which couples the
electron spin ${\bf S}$ and a nuclear spin ${\bf I}_i$:
\begin{eqnarray}
\hat{H} = \sum_i A_i \mid \psi \left( \bf{R}_i \right) \mid ^2
\left( \hat{S}_z \hat{I}_{i,z} + \hat{S}_+ \hat{I}_{i,-} + \hat{S}_-
\hat{I}_{i,+} \right),
\end{eqnarray}
where the sum goes over all nuclei in the QD electron localization
volume. The interaction strength of the electron with the $i-$th
nucleus is determined by the hyperfine constant $A_i$ and the
electron density $\mid \psi \left( \bf{R}_i \right) \mid ^2$ at
the nuclear site $\bf{R}_i$. $\hat{H}$ mediates processes in which
the spins of electron and nucleus are mutually flipped, as
described by the products of raising and lowering operators
$\hat{S}_\pm$ and $\hat{I}_{i,\pm}$, which increase and decrease
the spin projections $S_z$ and $I_{i,z}$ along the quantization
axis $z$, respectively. Two effects of the hyperfine interaction
on polarized electron spins are possible.

First, due to the limited number of nuclear spins interacting with
the electron spin in the QD, typically on the order of $N_L \sim
10^5$, a random correlation of the nuclear spins may create a
small nuclear polarization fluctuating from dot to dot. The
nuclear spin fluctuations (NSF) act on the electron spins as an
effective magnetic field, $B_f \propto 1/\sqrt{N_L}$, with random
magnitude and orientation~\cite{MerkulovPRB02}. The electron spin
precession about this field gives rise to a relatively fast
dephasing of the electron spins in a QD ensemble and, therefore,
to a decay of the electron spin polarization. Theoretical
estimates for GaAs QDs give dephasing times in the nanoseconds
range \cite{MerkulovPRB02,KhaetskiiPRL02} and predict a three-fold
decrease of the electron spin polarization in an ensemble due to
the precession. Experiments performed so far confirm a rapid
electron spin dephasing in an InAs QD ensemble due to the spin
precession in the NSF field~\cite{BraunPRL05,LombezPRB07}.
Thereafter, due to relatively slow flip-flop processes in the
nuclear spin system which change the NSF orientation on a
microsecond time scale, the electron spin orientation may be
further destroyed. This is assumed to be the most efficient
relaxation mechanism for electron spins in QDs
\cite{MerkulovPRB02,KhaetskiiPRL02} due to the suppression of
other mechanisms, in particular of the electron spin relaxation
via the spin-orbit interaction \cite{KhaetskiiPRB00}.  For
completeness we note that the hyperfine interaction is found to be
also the main relaxation mechanism of electron spins  at cryogenic
temperatures in n-doped bulk GaAs \cite{DzhioevPRL02} and
GaAs/(Al,Ga)As quantum wells \cite{DzhioevPRB02,Gerlovin06}.

The second effect of the hyperfine interaction appears when the
nuclear spins have been polarized somehow. The hyperfine
interaction with polarized nuclear spins affects the electron spin
like an effective magnetic field (the nuclear field), $B_{N}$,
causing a Zeeman splitting of electronic levels or an energy shift
of them (the Overhauser shift) in presence of an external magnetic
field. This effect has been observed in single-dot spectroscopy
studies
\cite{GammonPRL01,BrackerPRL05,YokoiPRB05,LaiPRL06,EblePRB06,%
BraunPRB06,MaletinskyPRB07,TartakovskiiPRL07,MaletinskyPRL07,%
UrbaszekPRB07}. In particular, an Overhauser shift of several tens
of $\mu$eV for GaAs interface QDs has been found
\cite{GammonPRL01} which corresponds to a 65\% polarization of the
nuclear spins and a nuclear field $B_N \sim 1.2$~T.

A nuclear spin polarization considerably affects also the electron
spin relaxation. First, when the electron spin is oriented
parallel to the nuclear field, the NSF effect is weakened because
the electron spin precesses about the total field whose
orientation is close to that of $B_N$ when $B_N \gg B_f$.
Second, the polarization of nuclear spins damps their flip-flop
processes~\cite{DengPRB06} and, correspondingly, suppresses
electron spin relaxation related to the NSF reorientation. Recent
experiments have demonstrated very long spin memories in
QDs~\cite{Elserman04,Kroutvar04,IkezawaPRB05,PalJPSJ06,OultonPRL07,GreilichScience07}
even though the role of the nuclear spin polarization in this
effect is still not totally clarified yet.

The nuclear spin dynamics may be also considerably modified by the
hyperfine interaction with polarized electron spins. The creation
of a dynamic nuclear polarization (DNP) via interaction with
optically polarized electrons is well known and has been
extensively discussed in literature (see, e.g.,
Ref.~\onlinecite{Zakharchenya}).

The DNP magnitude is usually measured by detecting the Overhauser
shift of optical transitions from (or to) electron Zeeman levels.
Due to the required high spectral resolution such measurements are
possible only on single QDs because of the large inhomogeneous
broadening of the optical transitions in a dot ensemble (typically
larger than 10~meV) compared to the Overhauser shift (a few or
several tens of $\mu$eV). Another method relies on studying the
photoluminescence polarization in a longitudinal or tilted
magnetic field, and has been successfully exploited for bulk
semiconductors (see in Ref.~[\onlinecite{Zakharchenya}] Chs.~2 and
5). However, it has still not been applied widely to QDs
\cite{PalPRB07}. To date, little is known about the DNP magnitude
in a QD ensemble and the role of DNP in the electron spin
relaxation \cite{OultonPRL07}. The NSF effect has also hardly been
studied because of the lack of sensitive experimental
methods~\cite{BraunPRL05}. In particular, nothing is known about
the effect of the electron spin polarization on the NSF.

Here we present results of detailed investigations of the
hyperfine interaction in singly negatively charged (In,Ga)As/GaAs
self-assembled QDs. Due to the presence of the resident electron
in a QD, its spin is an efficient tool for probing the nuclear
spin system. Exploiting two experimental methods, we have studied
the effects of NSF and DNP on the spin polarization of the
resident electrons in the QDs. In particular, we have obtained the
average magnitudes of the effective magnetic fields of NSF and DNP
and their dependence on the polarization and power density of the
optical excitation.

Recently we have studied the electron-nuclei interaction in a
regime in which we observed indications for a strong coupling
between the two systems, leading to the formation of a ``nuclear
spin polaron'' complex \cite{OultonPRL07}. These indications were
found under conditions of strong optical pumping realized by both
high optical excitation powers and long illumination times. Here
we address another regime, in which only rather short optical
exposure times were used. The goal of our study was to develop an
understanding of nuclear effects under these conditions, which
would correspond to the initial stages of nuclear polaron
formation.

\section{Experiment}

We studied two samples containing 20 layers of (In,Ga)As/GaAs
self-assembled QDs separated by 60 nm thick GaAs barriers.
Si-$\delta$ doped layers are incorporated 20 nm below each dot
layer. The as-grown InAs QD sample underwent a post-growth rapid
thermal annealing treatment for a duration of 30 s, either at a
temperature of 900$^{\circ}$C (sample $\# 1$) or 945$^{\circ}$C
(sample $\# 2$). This leads to diffusion of indium out of the
QDs~\cite{LangbeinPRB04}, and results in a shift of the
photoluminescence (PL) photon energy of the QD ground state from
$\sim 1.03$~eV for the as-grown sample to 1.338 and 1.396~eV for
the annealed samples, respectively. The average carrier
concentration in the QDs due to the barrier delta-doping is
approximately one electron per dot, as confirmed by Faraday
rotation spectroscopy \cite{GreilichPRL06}. We studied also a
quantum well heterostructure with two coupled 8~nm wide
In$_{0.09}$Ga$_{0.91}$As quantum wells separated by a thin, 1.7~nm
wide GaAs barrier was grown as a reference sample on an undoped
GaAs substrate by molecular beam epitaxy. It contains an n-doped
GaAs buffer layer which serves as source of electrons for the QWs.
The two-dimensional electron gas density in the QWs does not
exceed 10$^{10}$~cm$^{-2}$. The samples were immersed in pumped
liquid helium at temperature $T=$2~K.

We have explored the dynamics of the electron spin polarization
induced by circularly polarized pump pulses from a Ti:Sapphire
laser (pulse duration 1.5~ps at a pulse repetition frequency
75.6~MHz). Excitation with a fixed polarization helicity (either
$\sigma^+$ or $\sigma^-$) was used to study the DNP effect. When
we studied the NSF effect, the DNP was suppressed by modulation
of the circular polarization between $\sigma^+$ and $\sigma^-$
(hereafter referred to as $\sigma^+/\sigma^-$-excitation) at a
frequency of 25 or 50~kHz. A photo-elastic modulator or an
electro-optical modulator with sharp leading and falling edges
($<1$~$\mu$s) of the pulses followed by a quarter wave plate was
used for polarization control.

Two types of experiments were performed. First, we used resonant
optical excitation of the lowest optical QD transition and
measured the Faraday rotation signal (see details in Sec.~III).
Second, we measured the circular polarization of the PL, both
time-integrated and time-resolved, under non-resonant excitation
of the QDs. Time-resolved PL spectra were detected by a
synchro-scan streak camera attached to an 0.5-m spectrometer. The
time resolution was about 20~ps.

In the time-integrated measurements of the PL, we used special
timing protocols for modulation of the excitation intensity and
polarization. The intensity was modulated by an acousto-optical
modulator, which formed microsecond-long trains of picosecond
pulses. Four trains each of 7.5~$\mu$s duration with varying
polarizations formed a series which was repeated each 40~$\mu$s.
The primal polarization of the laser beam was $\sigma^+$. When we
studied the NSF effect, the polarization of the two last pulses
within a period was changed by an electro-optical modulator from
$\sigma^+$ to $\sigma^-$ so that the excitation was non-polarized
on average. This was done to avoid DNP in the QDs. In the study of
the DNP effect, the polarization was modulated non-symmetrically,
i.e., three pulses had one helicity and the fourth one had
opposite helicity, to obtain non-zero time-averaged circular
polarization of the excitation. The PL was detected for $\sigma^+$
polarization during the second and fourth pulses which were
$\sigma^+$- and $\sigma^-$-polarized in both cases, respectively.
The timing protocols therefore allowed us to study both nuclear
spin effects under almost identical experimental conditions. Note
that subsequent pulses in each train are separated by a 13.2-ns
time interval which is too short to change the electron or the
nuclear spin dynamics. Therefore a pulse train should have the
same impact as a pulse of the same duration as the train, and we
will denote in the following a pulse train just as long pulse.

All experiments were performed in a longitudinal magnetic field
oriented parallel to the structure growth axis and the light
wavevector (Faraday geometry). The field defines the spin
quantization axis, which we take as the $z$-axis [see Eq. (1)].
In this geometry, the components $B_{N,x}$ and $B_{N,y}$ of
the nuclear field normal to the external field are determined by
the NSF, while the component $B_{N,z}$ is determined by both the
NSF and DNP.

\section{NSF effect in Faraday rotation}

Faraday rotation (FR) spectroscopy has been widely exploited for
studying spin dynamics (see, e.g.,
Ref.~[\onlinecite{Spintronics}], Ch.~5). Using this technique,
sketched in Fig.~1(b), the circularly polarized pump beam orients
the electron spins along the optical axis and the linearly
polarized probe beam tests it. Detected is the rotation angle of
the probe polarization plane. The rotation occurs due to the
photo-induced Faraday effect which arises from an effective
magnetization created by the optically oriented electron spins in
the QDs leading to circular birefringence of the sample. We used a
degenerate pump-probe method in which the wavelengths of pump and
probe are the same.

\begin{figure}[hbt]
\includegraphics*[width=7.5cm]{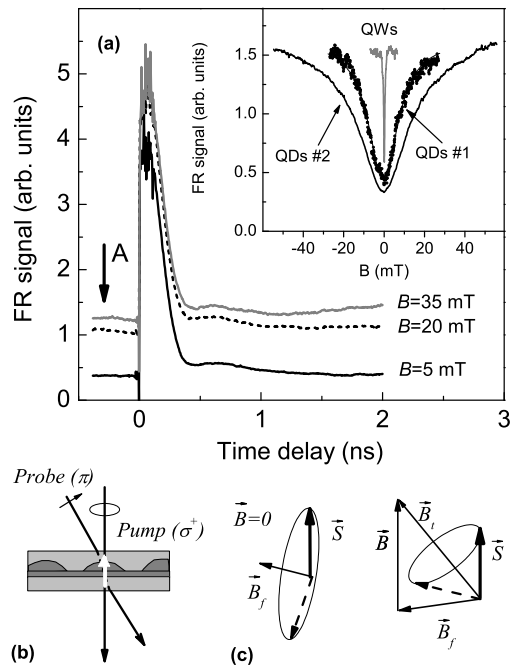}
\caption[] {(a) FR signal of sample $\# 2$ for different
longitudinal magnetic field strengths, as noted at each curve.
Sample temperature $T=2$~K; pump power density $P=20$~W/cm$^2$.
The arrow A shows the time delay at which the magnetic field
dependence of the signal amplitude was measured. Inset: field
dependence of the FR signal for different studied samples. (b)
Schematic illustration of the polarization resolved pump-probe
method for measuring Faraday rotation (see text). (c) Mechanism of
suppression of the NSF effect by an external magnetic field. }
\label{fig:1}       
\end{figure}

We have studied the FR signal of the QD samples in dependence of
the longitudinal magnetic field. Typical traces of the signal
versus the delay between pump and probe are shown in Fig.~1(a).
They were measured for $\sigma^+$ / $\sigma^-$-excitation for
which no DNP occurs. Each trace consists of a rapidly decaying and
a long-lived component. The fast component with a decay time of
about 0.3~ns is related to recombination of the electron-hole
pairs created by the strong pump pulse. The long-lived component
is due to spin polarization of the resident electrons
\cite{GreilichPRL06,GreilichScience}.

Application of a relatively small magnetic field causes a strong
increase of the amplitude of the long-lived FR component so that
it becomes pronounced even at negative delays where it originates
from the preceding laser pulses. We measured the field dependence
of the FR signal at a fixed negative delay time indicated by the
arrow. This delay is $\sim 13$-nanoseconds after the preceding
pulse, at which all processes such as electron-hole recombination,
have decayed and the signal is determined solely by the spin
orientation of the resident electrons. The inset in Fig.~1(a)
shows the magnetic field dependence of the FR signal which reveals
a dip around $B=0$. Its full width at half maximum (FWHM) varies
considerably for the different studied samples.

The dip in the field dependence can be explained by the model of
Merkulov {\em et al.} \cite{MerkulovPRB02} which is illustrated in
Fig.~1(c). The electron spin $\vec{S}$ is initially oriented by
laser excitation along the optical excitation axis which coincides
with the growth direction of the heterostructure. In presence of
the NSF and an external magnetic field $B$, the electron spin
rapidly precesses about the total field, $\mathbf{B}_t =
\mathbf{B}_{f}+\mathbf{B}$ with a frequency $\omega=|g_e| \mu_B
B_t/\hbar$ where $|g_e| \approx 0.5$ is the QD electron $g$-factor
and $\mu_B$ is the Bohr magneton. This precession results in an
oscillation of the spin projection onto the $z$-axis for each QD.
Because of the spread of $B_f$ in the dot ensemble, the
oscillations in different dots occur with different frequencies.
Destructive interference of the oscillations gives rise to a decay
of the $z$-projection of the average electron spin in the
ensemble. The FR signal is proportional to this $z$-projection of
the spin and, therefore, is expected to decay too. For zero
external magnetic field, the decay time may be estimated using a
Gaussian distribution for $B_f$ [see
Ref.~\onlinecite{MerkulovPRB02}]: $\tau_{NSF}=1 / \omega \sim
1$~ns, where we used a value $B_f = 15$~mT as estimated below.

In Fig.~1(a) we do not observe a decay with this characteristic
time. On the contrary, a long-lasting signal is observed which
seems to be in contradiction with the prediction. This discrepancy
can be explained as follows.

The electron spin polarization is not very efficient for the
experimental conditions used in the Faraday rotation experiment.
The low polarization efficiency is related to a high stability of
the spins of the resonantly photo-created holes which persists
during their lifetime (and may go up to tens of nanoseconds for
resident holes~\cite{LaurentPRL05}). As a result, the holes
recombine predominantly with the photo-created electrons and
almost no spin polarization is transferred to the resident
electrons. That is why the one-nanosecond decay of the
polarization cannot be resolved. Still, because this small
electron spin polarization lives for a long time in the
microseconds range for the QDs under study~\cite{GreilichScience},
a long-lasting contribution can build up during the many pump
pulses within the signal recording time, so that it becomes
observable and is almost constant in the scanned range of delays.

The magnitude of the spin $z$-projection depends on the ratio of
$B$ and $B_f$. It decreases with decreasing external magnetic
field and falls to $1/3$ of its initial value $S_0$ (determined by
the optical excitation) at $B=0$ corresponding to $B_t=B_f$, as it
becomes randomly distributed over all directions
\cite{MerkulovPRB02}. In the other limit of strong $B \gg B_f$,
$B_t$ is mostly directed along the $z$-axis and the spin
projection is retained up to its initial value $S_0$. The
theoretical magnetic field dependence of the electron spin
polarization obtained by Merkulov {\it et
al.}~\cite{MerkulovPRB02} for a Gaussian statistics of the NSF can
be approximated by the form~\cite{PetrovPRB08}:
\begin{equation}
S(B)=S_m\left(1-\frac{A_{f}}{1+\left(B/B_f\right)^2}\right),
\label{eqn1}
\end{equation}
where $A_{f}$ is the amplitude of the dip and $S_m$ is the spin
polarization at large magnetic fields. We use this function for
quantitative analysis of our experimental data.

The FWHM of the dip may serve as a measure of the effective
magnetic field of the NSF. As seen from the inset in Fig.~1(a),
$B_f$ is about 10~mT and 15~mT for QD samples $\# 1$ and $\# 2$,
respectively~\cite{Note-Width}. Theoretical estimates for GaAs QDs
with $10^{5}$ nuclei \cite{MerkulovPRB02,KhaetskiiPRL02} give very
similar values. The fluctuation field $B_f$ is much smaller (about
0.3~mT) for the reference (In,Ga)As/GaAs QWs. The strength of
$B_f$ depends on the electron localization volume, i.e., on the
number $N_L$ of nuclei at which the electron density has a notable
magnitude: $B_f \propto 1/\sqrt{N_L}$. In quantum wells, the
electron is only weakly localized by interface fluctuations along
the two in-plane directions. Therefore its localization volume is
large compared to the QD case and $B_f$ is small. This is in
accord with our experimental findings. From the $B_f$-ratio of
$\sim 70$ for the QDs and QWs under study we conclude that the
electron-localization volume in the wells is about three orders of
magnitude larger than that in the dots.

The Faraday rotation technique is restricted, however, by several
limitations for studying nuclear spin effects. As described, under
the used low field conditions the optical pumping of the electron
spin which we could achieve was quite small. Correspondingly, also
the magnitude of the DNP under optical excitation with fixed
helicity was quite small. Further, the FR signal does not have a
natural scale and, therefore, it is unclear what degree of
electron spin polarization is achieved in these experiments.
Finally, the background signal related to scattered light of the
pump beam prevents detection of the spin depolarization by the NSF
with high accuracy.

\section{NSF and DNP in PL polarization}

To study the effects of NSF and DNP more quantitatively, we used
another experimental method, namely we exploited the effect of
negative circular polarization (NCP) of the PL which appears for
the studied QDs under certain experimental conditions. We excited
the QDs quasi-resonantly (via intra-dot optical transitions) or to
the low-energy wing of the wetting layer absorption band at energy
$E_{exc}=1.467$~eV. We found that there is no principal difference
in the PL polarization for these two excitation conditions. The PL
reveals circular polarization under circularly polarized
excitation [see Fig.~2(a) and (b)]. The polarization degree has
been calculated by: $\rho_c = (I^{++}-I^{+-})/(I^{++}+I^{+-})$,
where $I^{++}$ ($I^{+-}$) is the PL intensity for co- (cross-)
polarization of excitation and detection. The sign of polarization
for the lowest QD optical transition is negative, so that the PL
intensity is stronger in cross- than in co-polarization. This is
confirmed by the PL kinetics which has been measured at the energy
of the PL band maximum [Fig.~2(c) and (d)]. For the further study
of the NCP effect, we used excitation at an energy
$E_{exc}=1.467$~eV because excitation at lower energy requires
much stronger pumping powers due to the smaller absorption of the
intra-dot optical transitions.

\begin{figure}[hbt]
\includegraphics*[width=8.5cm]{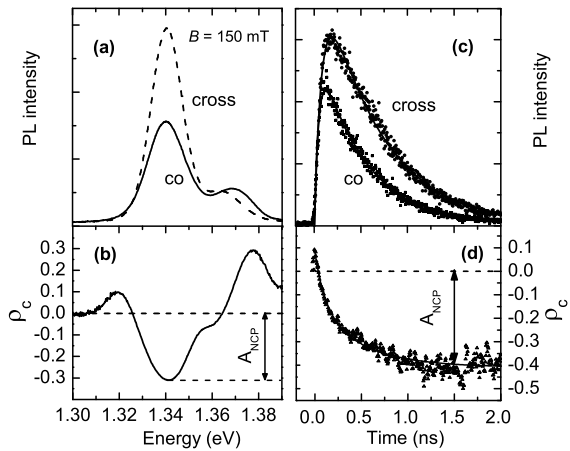}
\caption[] {(a) PL spectrum of sample $\# 1$ for co- ($\sigma^+
\sigma^+$, solid curve) and cross- ($\sigma^+ \sigma^-$, dashed
curve) polarization of excitation and detection. Excitation at
$E=1.467$~eV with a power density $P=20$~W/cm$^2$ for $B=150$~mT;
$T=2$~K. (b) Spectrum of the circular polarization degree of the
PL resulting from (a). (c) PL kinetics measured at
$E_{PL}=1.340$~eV at $B=150$~mT in the two polarization
configurations. (d) Polarization kinetics of the PL resulting from
(c).}
\label{fig:2}       
\end{figure}

NCP of the ground state PL has been reported earlier for
InAs~\cite{CortezPRL02,LaurentE04,LaurentPRB06,OultonPRL07},
InP~\cite{DzhioevFTT98,IkezawaPRB05}, and GaAs~\cite{WarePRL05}
QDs. Although different mechanisms of NCP formation for singly
negatively charged QDs have been discussed~\cite{DzhioevFTT98,CortezPRL02,%
WarePRL05,LaurentPRB06,IkezawaPRB05}, it is commonly accepted that
the large value of the NCP observed under strong optical
excitation is due to accumulation of spin polarization by the
resident electrons. The mechanism of the electron spin orientation
is related to the spin relaxation of the photocreated holes in the
QDs. These depolarized holes recombine with non-polarized resident
electrons. Correspondingly, polarized electrons provided by the
optical excitation are accumulated in the QDs.

The hole spin is quite stable when the hole is in the QD ground
state ~\cite{LaurentPRL05}. Therefore the efficiency of spin
polarization of the resident electrons is not high for optical
excitation far below the wetting layer transition. Probably this
is the reason why we did not observe an NCP exceeding 20\% for
intra-dot excitation with an energy $E_{exc} < 1.43$~eV.

The mechanism of NCP for high-energy excitation, when free
electrons and holes are created in the barrier layers, may be
related to dark exciton formation~\cite{DzhioevFTT98}. However,
for quasi-resonant excitation when dark excitons cannot be
created, the electron-hole spin flip-flop process plays the main
role for
NCP~\cite{CortezPRL02,KavokinPSS03,IkezawaPRB05,WarePRL05}. In our
subsequent analysis, we consider this process to be responsible
for the NCP observed in our experiments.

The detailed NCP formation via the flip-flop process goes as
follows. If the resident electron is polarized, a photocreated
electron with the same spin polarization cannot relax to the
ground state because of Pauli principle unless flipping its spin.
The flip is most efficient if accompanied by a simultaneous flop
of the hole spin. As a result of this flip-flop process, the
electron spins become paired to a singlet and the spin of the hole
is inverted. Recombination of the hole with one of the electrons
gives rise to emission of a $\sigma^+$ ($\sigma^-$) polarized
photon for $\sigma^-$ ($\sigma^+$) polarization of the excitation.
This corresponds to a negative sign of polarization. The analysis
also shows \cite{IkezawaPRB05} that the amplitude of the NCP,
$A_{NCP}$, extracted from the PL kinetics may be used as a measure
of the spin polarization of the resident electron [see Fig.~2(d)].
In particular, for high enough spin polarization, the approximate
relation
\begin{equation}
A_{NCP} \approx 2 \left< S_z \right> \label{eqn2}
\end{equation}
is valid. The factor ``2'' appears because the theoretical upper
limit $A_{NCP}=1$ for NCP, while for the electron spin projection
onto the optical axis of excitation, $\left< S_z \right> = 1/2$.
We will use relation (\ref{eqn2}) in our subsequent analysis of
the experimental data.

The amplitude of the NCP obtained in time integrated PL at the
maximum of the emission band is slightly smaller than the one
obtained from PL kinetics [compare Fig.~2(b) and (d)] due to
averaging the polarization over the PL pulse. In this work, we
studied the behavior of $A_{NCP}$ mainly in time integrated PL
because of the related simplicity of measuring the magnetic field
dependence of $A_{NCP}$ through scanning the field strength. We
neglect the small difference between the $A_{NCP}$ values obtained
in time resolved and time integrated experiments and consider the
time integrated value as a quantitative characteristic of the spin
polarization of the resident electrons.

We measured the magnetic field dependence of $A_{NCP}$ at
different pump-power densities of optical excitation. Two regimes
of circular polarization modulation of the excitation have been
used. In the first one, the polarization was rapidly switched from
$\sigma^+$ to $\sigma^-$. In this case, DNP did not build up which
allowed us to study the NSF effect only. For DNP-studies
excitation with predominate polarization $\sigma^+$ or $\sigma^-$
was used. A representative set of the obtained results for sample
\#1 is given in Fig.~3.

\begin{figure}[hbt]
\includegraphics*[width=8.5cm]{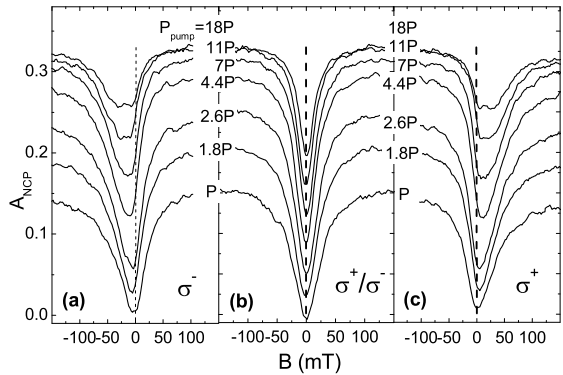}
\caption[] {Magnetic field dependence of the NCP amplitude for (a)
$\sigma^-$, (b) $\sigma^+$/$\sigma^-$, and (c) $\sigma^+$
polarization of the excitation at different pump-power densities
given at each curve. Sample \#1. $T=2$~K. $P=P_0=1$~W/cm$^2$.}
\label{fig:3}       
\end{figure}

\subsection{Nuclear spin fluctuations}

First we discuss the NSF effect [Fig.~3(b)]. The magnetic field
dependence of the PL polarization is seen to reveal a dip around
$B=0$ which is very similar to the one observed in the Faraday
rotation experiment. Therefore we treat the field dependence of
the PL polarization also in the frame of the model used above for
description of the FR experiments, even though the picture may be
more complex under non-resonant PL excitation due to the energy
relaxation processes. A theoretical simulation of the NSF effect
in annealed InAs QDs~\cite{PetrovPRB08} gives rise to a similar
dip in the electron spin polarization with a FWHM which is close
to the one observed at low pumping. This result further supports
our interpretation of the dip.

Two important effects are seen with increasing pump power. First,
the FWHM of the dip {\it decreases} from 30~mT at 1 W/cm$^2$ to
15~mT at 30~W/cm$^2$. Second, the amplitude of the dip $A_f$
normalized to the PL polarization magnitude, $A_m$, measured at $B
\gg B_f$ also decreases with pump power [see Fig.~4(c)]. These
dependencies are shown in Figs.~4(a) and (b). Both effects point
out that the influence of the NSF in the electron spin
depolarization becomes weaker when optical pumping becomes
stronger even when no dynamic nuclear polarization appears due to
modulation of the excitation polarization. We discuss these
effects quantitatively in the next section.

\begin{figure}[hbt]
\includegraphics*[width=8.5cm]{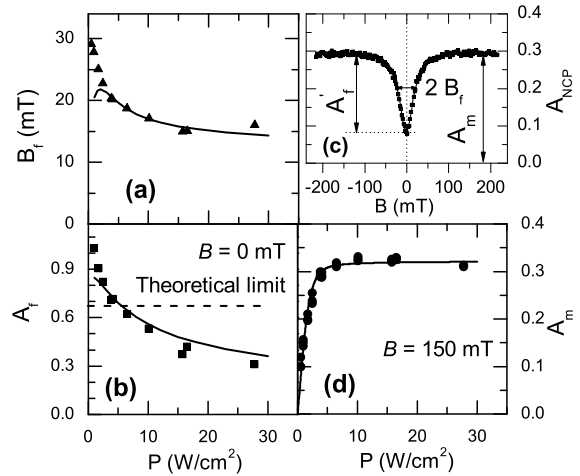}
\caption[] {Pump power dependencies of the effective magnetic
field of the NSF, $B_f$, (a) and of the normalized amplitude of
the spin polarization dip, $A_f=A_f^{'}/A_m$, (b) for sample \#1
~\cite{Note-ampl}. The ``theoretical limit'' line shows the
dip depth, $A_f = 2/3$, expected for ``frozen'' NSF with spatially
isotropic distributions \cite{MerkulovPRB02}. Solid lines are
calculated in the frame of the ``melted'' NSF model using
Eqs.~(\protect{\ref{eqn4},\ref{eqn11},\ref{eqn12}}) with the
parameters $\Delta_z=15$~mT, $\tau_{e0}=3$~$\mu$s,
$\tau_{em}=60$~$\mu$s. Panel (c) shows a typical dip in the NCP
amplitude used to explain the notations. (d) Pump power dependence
of the NCP amplitude at $B=150$~mT. The solid line is calculated
using Eqs.~(\protect{\ref{eqn2},\ref{eqn11}}) with the parameters
$\left\langle S_{zf} \right \rangle = S_m = 0.16$ and
$T_{exc}^0=10$~$\mu$s. }
\label{fig:4}       
\end{figure}

\subsubsection{``Frozen'' nuclear spin fluctuations}
To understand the observed behavior of the electron spin
polarization, we start by considering the effect of ``frozen''
nuclear spin fluctuations. For this purpose, we use a simple model
based on the approach developed by Merkulov {\em et
al.}~\cite{MerkulovPRB02} We assume that the statistics of the
$\alpha$-component ($\alpha = x$, $y$, and $z$) of the NSF field can
be described by a Gaussian function for the probability density
distribution:
\begin{equation}
W(B_{f\alpha}) = \frac{1}{\sqrt{\pi}\Delta_\alpha}
\exp\left[-\frac{(B_{f\alpha})^2}{\Delta_\alpha^2}\right].
\label{eqn3}
\end{equation}
Here $\Delta_\alpha$ is the dispersion of the $B_{f\alpha}$. We
emphasize that the observation of a drop of the NSF field $B_f$
with increasing pump power in Fig. 4(a) cannot be explained by
assuming that these dispersions stay isotropic under optical
pumping of the nuclear spins. Therefore, in contrast to
Ref.~\onlinecite{MerkulovPRB02}, we suggest that the distributions
become unequal when pumping the nuclear spin system by
alternatingly polarized electron spins: $\Delta_x = \Delta_y \ne
\Delta_z$. With this assumption we have calculated the
$z$-component of the ensemble-averaged electron spin polarization
in the field of ``frozen'' NSF as function of the external
magnetic field:
\begin{equation}
\left< S_{zf} \right> = 2\pi \int\limits_{-\infty}^{\infty}
W(B_{fz})d B_{fz} \int\limits_0^{\infty} W(B_{\rho}) S_z(B)
B_{\rho} dB_{\rho}. \label{eqn4}
\end{equation}
Here we assumed a cylindrical symmetry of the nuclear spin
distribution and defined the radial field $B_{\rho} =
\left[B_{fx}^2 + B_{fy}^2\right]^{1/2}$ with a distribution
$W(B_{\rho})=W(B_{fx})W(B_{fy})$. $S_z(B)$ is the time-averaged
$z$-component of the electron spin polarization reduced relative
to its initial value $S_m$ due to its fast precession about the
total magnetic field, ${\bf B}_t={\bf B}_f+{\bf B}$, for each
particular configuration of the fluctuating nuclear field ${\bf
B}_f$ [see Fig.~1(c)]:
\begin{equation}
S_z(B) = S_m [\cos{\gamma(B)}]^2 =
\frac{(B_{fz}+B)^2}{(B_{fz}+B)^2+B_{\rho}^2} \label{eqn5},
\end{equation}
where $\gamma(B)$ is the angle between ${\bf B}_t$ and the initial
electron spin orientation. For the averaging in Eq.(\ref{eqn4}) we
used distribution functions $W \left(B_{f\alpha}\right)$ with
equal excitation power dependent half widths for the transverse
NSF components:
\begin{equation}
\Delta_y=\Delta_x=\frac{\Delta_x^0}{\sqrt{1+kP}}, \label{eqn7}
\end{equation}
where $k$ is a scaling factor for the excitation power $P$. The
power dependence (\ref{eqn7}) has been determined in the model
described in Appendix A.

The results of the calculations for the magnetic field dependence
of the electron spin polarization at different excitation powers
are shown in Fig.~5(a). The dip profile is well approximated by
eq. (1) if we use the parameter $B_f=1.39 \Delta_z$ at low
excitation powers when $\Delta_x=\Delta_z$. In the calculations of
the power dependence, we use a value $k=2/3$~cm$^2$/W to obtain
the approximately two-fold decrease of the dip at $P =
30$~W/cm$^2$, as observed experimentally [see Fig.~4(b)]. The dip
FWHM and amplitude are shown in Fig.~5(c) and (d) by the dashed
lines. The decrease of the dip amplitude observed at strong
pumping can be satisfactory explained by the decrease of the
transverse component of the NSF.

\begin{figure}[hbt]
\includegraphics*[width=8.5cm]{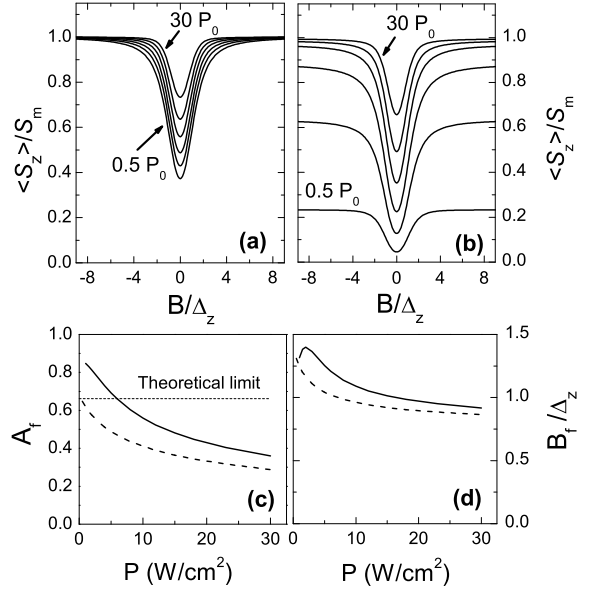}
\caption[] {(a) Magnetic field dependence of $\left< S_z \right>$
(normalized to $S_m$) calculated using the model of ``frozen'' NSF
for different excitation densities: $P=0.5$, 1.5, 3, 6, 12,
30~W/cm$^2$. (b) The same for ``melted'' NSF. (c) Dependence of
the dip amplitude, $A_f$, on the power density for ``melted''
(solid line) and ``frozen'' (dashed line) NSF. (d) The same for
the effective magnetic field, $B_f$. }
\label{fig:5}       
\end{figure}

\subsubsection{``Melted'' nuclear spin fluctuations}

The model considered above allows us to describe the tendency in
the dip behavior at relatively high excitation powers. However,
for weak excitation when the process of optical spin polarization
is slow, the model of ``frozen'' NSF cannot describe the increase
of the dip amplitude up to unity [see Fig.~4(b)] as well as the
small magnitude of the spin polarization even at large magnetic
fields [Fig.~4(d)].

To describe this limit, we consider a slow variation of the NSF in
time as discussed by Merkulov {\it et al.} \cite{MerkulovPRB02}.
These ``melted'' NSF may further depolarize the electron spin
during a characteristic time in the microseconds range.
Depolarization can occur only when the intrinsic electron spin
lifetime (i.e., the one not related to the NSF effect) is large
enough. Studies of the spin coherence in our QDs in presence of a
transverse external magnetic field, when the NCP effect is
suppressed, show a characteristic spin relaxation time of the
resident electrons in the microseconds
range~\cite{GreilichPRL06,GreilichScience,CherbuninNANO07}.

For determining the spin polarization along the $z$-axis, we have
solved a rate equation model for the populations of the spin-split
electron states. Besides we have taken into account that the spin
polarization is measured in our experiments in the second half of
a long pulse from $\tau_p/2$ to $\tau_p$. Details are described in
Appendix A. The resulting spin polarization is given by:
\begin{eqnarray}
\left\langle S_z \right \rangle &=& \left\langle S_{zf}\right\rangle %
\Bigl[\frac{2}{\gamma \tau_p}\left(\Delta n_0 + \frac{\gamma_{exc}}{\gamma}\right) \nonumber\\
&\times&\left(\exp(-\gamma \tau_p/2)-\exp(-\gamma
\tau_p)\right)-\frac{\gamma_{exc}}{\gamma}\Bigr], \label{eqn11}
\end{eqnarray}
where $\left\langle S_{zf}\right\rangle$ is the spin polarization
conserved after action of the ``frozen'' NSF, given by
Eq.~(\ref{eqn4}). $\gamma = \gamma_e + \gamma_{exc}$, with the
excitation rate $\gamma_{exc}=\gamma_{exc}^0 P$, where
$\gamma_{exc}^0$ is the rate at $P_0=1$~W/cm$^2$, and $\gamma_e$
is the electron spin relaxation rate due to the time variations of
the NSF. $\Delta n_0$ is the population difference between the two
spin-split levels right before the pulse. This population
difference was created by the preceding pulse with opposite
circular polarization, which has partially relaxed during the dark
time between pulses.

Above all, we discuss the pump power dependence of the electron
spin polarization measured in an external magnetic field
sufficient for suppressing the NSF effect [Fig.~4(d)]. The
polarization saturates at a level of about 30\% of its maximal
value for excitation densities $P>5$~W/cm$^2$. This value is
considerably smaller than typically observed in single dot
spectroscopy~\cite{EblePRB06}. We attribute this difference mainly
to the coexistence of neutral, singly charged and multiply charged
dots in the QD ensemble~\cite{LaurentPRL05}. The PL of the neutral
QDs is linearly polarized or unpolarized because of the
anisotropic exchange interaction of electrons and holes in the
QDs. The PL polarization of the multiply charged QDs is reduced
too~\cite{IkezawaPRB05,LaurentPRB06}. The contribution of these
dots reduces the total PL polarization of the ensemble, as only
the singly charged dots provide a high polarization of the
emission.

We use Eq.~(\ref{eqn11}) to analyze the power dependence in
Fig.~4(d) quantitatively. For this purpose, we use an electron
spin relaxation rate, $\gamma_e=\gamma_{em} \approx
1/6$~$\mu$s$^{-1}$, as measured in
Ref.~\onlinecite{CherbuninNANO07} for this sample in presence of a
magnetic field suppressing the NSF, and consider quantity
$\gamma_{exc}^0$ as
fitting parameter. As seen from Fig.~4(d), the fit satisfactorily
reproduces the experiment. The obtained value of the fitting
parameter $\gamma_{exc}^0 = 0.1$~$\mu$s$^{-1}$ means that the spin
orientation of the resident electron is on average restored during
a time interval $T_{exc}^0=(\gamma_{exc}^0)^{-1}=10$~$\mu$s by
optical excitation with a power density $P_0=1$~W/cm$^2$.

This value can be compared with a simple estimate based on the
experimental conditions and the structural parameters of our
sample. Taking into account the total QD density in the 20 layers,
$\rho_{QD} \sim 4 \times 10^{11}$~cm$^{-2}$ [see
Ref.~\onlinecite{GreilichPRL06}], we estimate that about 100
photons hit each dot during $T_{exc}^0$ at $P_0=1$~W/cm$^2$.
Further, only a small fraction, $a$, of these photons is absorbed
at the excitation photon energy $\hbar \omega = 1.467$~eV. Our
measurements show that the probability of absorption is about 20
\%. Finally, the efficiency of electron spin polarization, $q$,
most likely is relatively small [e.g., $q=0.05$ for InP
QDs~\cite{IkezawaPRB05}]. From these estimates it is reasonable
that the product $a \cdot q$ can be as small as 0.01 and therefore
only one photon out of 100 impinging photons gives rise to
polarization of a resident electron spin during $T_{exc}^0$.

Let us consider now the magnetic field dependence of the spin
polarization at low excitation density. In absence of a magnetic
field, the effective time $\tau_{e0}=(\gamma_{e0})^{-1}$ of electron
spin relaxation due to interaction with the ``melted'' NSF is on the
order of a few microseconds~\cite{MerkulovPRB02,PalPRB07}.
Application of a magnetic field suppresses the effect of the
``melted'' NSF and, correspondingly, slows down the electron spin
relaxation \cite{PalPRB07}. At present, there is no quantitative
theoretical description of the field dependence of the electron spin
relaxation rate. Therefore we use a phenomenological dependence
based on the experimental data by Pal {\em et al.}~\cite{PalPRB07}:
\begin{equation}
\gamma_e(B)=\gamma_{em}+\frac{\gamma_{e0}}{1+\left(B/B_f\right)^2}.
\label{eqn12}
\end{equation}
The second term on the right hand side describes the suppression
of spin transfer to the nuclear spin system when the angle between
the electron spin and the total magnetic field [see Fig.~1(c)]
becomes small and, correspondingly, the spin precession is
inefficient. In the following calculations, $\gamma_{e0}$ is
considered as fitting parameter.

We use Eqs.~(\ref{eqn4},\ref{eqn11},\ref{eqn12}) to determine the
magnetic field dependence of $\left\langle S_z \right \rangle$.
The results of these model calculations are shown in
Figs.~5(b)-(d) by the solid lines. In Fig.~4(a) and (b), the
calculated dependencies for the depth and halfwidth of the dip are
compared with the experiment. As seen, the dependences of $A_f$
and $B_f$ are well reproduced in the main range of pump densities
except for small values $P<3$~W/cm$^2$.

So, the presented analysis allows us to explain phenomenologically
the main features of the electron spin polarization in presence of
the NSF. The main assumption we have used to explain the
suppression of the NSF effect under strong pumping is the
reduction of the transverse NSF component relative to the
longitudinal one, for which the possible physical reason will be
discussed by the end of the next section. Consideration of the
time variations of the NSF (``melted'' NSF) allowed us to explain
semi-quantitatively the dip width and amplitude also in the limit
of low excitation densities [see Fig.~4(a) and (b)]. There is,
however, some disagreement between theory and experiment at this
low excitation. One of the possible reasons is that relation (2)
becomes invalid when the average spin polarization of the resident
electrons is small~\cite{IkezawaPRB05}. We also consider
Eq.~(\ref{eqn12}) to be oversimplified to describe correctly the
electron-nuclear spin dynamics in this case. Further theoretical
and experimental investigations are needed to clarify this
problem.

\subsection{Dynamic nuclear polarization}

The observed considerable suppression of the NSF transverse
components may hint a strong polarization of the nuclear spins
under the applied experimental conditions. To check this, we have
studied the DNP maintaining the experimental conditions used in
the NSF study except for the modulation of the optical
polarization. Usually DNP is studied for a fixed polarization of
optical excitation~\cite{Zakharchenya}. We, however, still used
partial modulation of the polarization to avoid modifications of
the optical setup. We used a relatively long pulse ($\tau_p =
22.5$~$\mu$s) of one polarization ($\sigma^+$ or $\sigma^-$) and a
three times shorter pulse (one pulse train) of opposite
polarization ($\sigma^-$ or $\sigma^+$, correspondingly) with the
same peak intensity. The dark time between the pulses was 2~$\mu$s
and the repetition frequency of the pulses was 25~kHz. This
excitation protocol allowed us to obtain a time-averaged net
polarization of one helicity and, correspondingly, to polarize the
nuclei.

As seen from Fig.~3(a) and (c), such excitation results in a shift
of the dip position away from zero magnetic field. This effect is
shown in more detail in Fig.~6  for an excitation power density
$P=16$~W/cm$^2$. The sign of the shift depends on the helicity of
the light polarization and the shift increases with pump power
density. Both observations are explained by DNP.

\begin{figure}[hbt]
\includegraphics*[width=8cm]{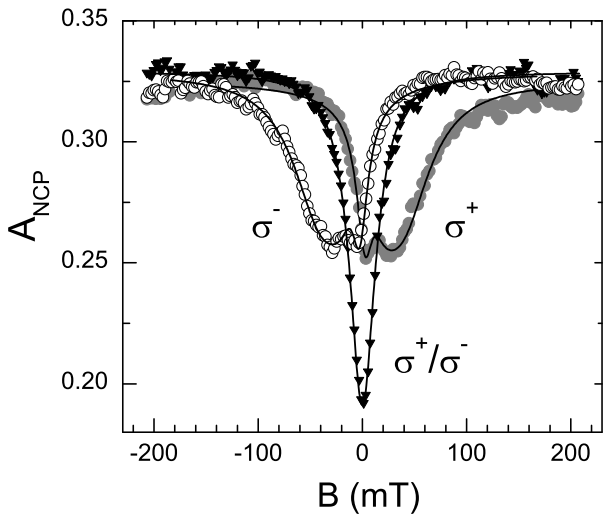}
\caption[] {Magnetic field dependence of the electron spin
orientation for sample \#1 under $\sigma^-$ (open circles),
$\sigma^+$ (filled circles) and $\sigma^+/\sigma^-$ (triangles)
polarized excitation with $P_{exc}=16$~W/cm$^2$. Lines are fits by
Eqs.~(\protect{\ref{eqn13},\ref{eqn14}}).
}
\label{fig:6}       
\end{figure}

The effective magnetic field of the DNP, $B_N$, is directed along
the external magnetic field or along the opposite direction,
depending on the pump helicity, so that the total magnetic field,
$B_t=B+B_N$, increases or decreases relative to the external
magnetic field, $B$. The magnetic field dependence of the PL
polarization is described by an equation similar to
Eq.~(\ref{eqn1}):
\begin{equation}
A_{NCP}(B)=A_m\left(1-\frac{A_{f}}{1+\left[(B+B_N)/B_f\right]^2}\right).
\label{eqn13}
\end{equation}
The dip in the electron spin orientation due to the NSF is still
present as long as a total orientation of the nuclear spins has
not been achieved, and should occur at $B_t=0$ when $B_N = -B$.
Thus the value of the dip shift may serve as a measure for the DNP
effective magnetic field. The increase of the dip shift with power
is inherently explained by the increase of the DNP.

Fig.~6 shows one more narrow dip in the PL polarization which
appears at nearly zero magnetic field for $\sigma^+$ and
$\sigma^-$ polarized excitation. The origin of this dip is related
to destruction of the DNP by dipole-dipole interaction between the
nuclear spins when the external magnetic field becomes smaller
than the local field, $B_L$, acting on a nuclear spin through its
neighbors. This effect gives rise to the following dependence of
the DNP on the magnetic field~\cite{Zakharchenya}:
\begin{equation}
B_N(B) = \frac{B_{Nm}}{1+\left[B_L/(B+B_e)\right]^2},
\label{eqn14}
\end{equation}
where $B_{Nm}$ is the maximal nuclear magnetic field reached for
fixed circular excitation at $B \gg B_L$.

In Equation~(\ref{eqn14}), we also took into account the effective
magnetic field of the electron at a nuclear site (Knight field),
$B_e$. We will not discuss this effect in detail~\cite{NoteKnight}
and consider $B_L$ and $B_e$ as fitting parameters. $B_e$
determines the shift of the narrow dip away from zero magnetic
field. Its value can reach several milliTesla for the strongly
localized electrons in QDs~\cite{LaiPRL06}. Our fit gives $B_e
\sim 10$~mT. The parameter $B_L$ characterizes the dip width whose
value is much larger in our case than the typical local field
(fraction of milliTesla~\cite{NoteKnight}). We assume that the dip
width increase is mainly due to the spread of the QD parameters,
in particular the Knight field, in the ensemble.

Examples of fitting the PL polarization by
Eqs.~(\ref{eqn13},\ref{eqn14}) are shown in Fig.~6. The fit
reproduces the magnetic field dependence of
$A_{NCP}$~\cite{Note-field} well which allows us to reliably
determine the effective magnetic field of the DNP. Its dependence on
the excitation power density is shown in Fig.~7.

\begin{figure}[hbt]
\includegraphics*[width=8cm]{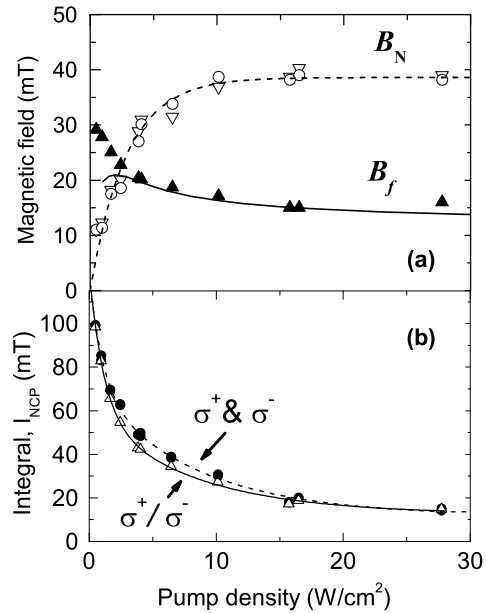}
\caption[] {(a) Effective magnetic field of the DNP, $B_N$, as
function of pump density. Circles and triangles are the experimental
data for $\sigma^-$ and $\sigma^+$ polarized excitation,
respectively; the dashed line is a guide to the eye. Data for the
effective magnetic field of the NSF $B_f$ from Fig.~4(a) are shown
for comparison. (b) Integral of the dip as function of the pump
density calculated by Eq.~(\ref{eqn25}) for $\sigma^+$- and
$\sigma^-$-polarized excitation (circles), as well as for
$\sigma^+/\sigma^-$-polarized excitation (triangles). Solid lines
are guides to the eye. Data have been taken for sample \#1. }
\label{fig:7}       
\end{figure}

$B_N$ rapidly rises with excitation power in the same range in
which the NSF field decreases and the electron spin orientation
increases [compare with Fig.~4(d)]. The DNP field saturates at a
value of about 40~mT at strong pumping. This value is small
relative to the theoretical limit of a few Tesla when all the
nuclei have been polarized~\cite{Zakharchenya}.

Simultaneously the NSF value does not decrease considerably any
further. To demonstrate this more clearly, we have calculated the
integral of the dip area in the electron spin polarization by
taking the difference between the line $y=A_m$ and the function
$y=A_{NCP}$ normalized to $A_m$, [see Fig~4(c)]:
\begin{equation}
I_{NCP}=\int{\left(1-A_{NCP}/A_m\right)\,dB} \label{eqn25}
\end{equation}
We consider this quantity as a measure of the effect of the
fluctuating nuclear field on the electron spin polarization. As
seen from Fig~7(b), the integral $I_{NCP}$ and correspondingly the
NSF decrease strongly with increasing pumping for all used
excitation polarization configurations.

\section{Discussion}

It is well established now that a significant polarization of the
nuclei takes a long time of optical pumping in the seconds or
minutes range~\cite{GammonPRL01,Zakharchenya,MakhoninPRL08} and
the nuclear polarization degree rises highly nonlinearly with
time~\cite{MaletinskyPRL07,Auer}. Due to the partially modulated
excitation polarization with a rather short period $T =
40$~$\mu$s (see Sect.~II) in our experiments, we effectively
study the early stage of the nuclear polarization dynamics.
Experimental studies~\cite{Auer} have confirmed that a $B_N$ with
a magnitude of a few tens of milliTesla is established on a
sub-millisecond time scale at strong enough pumping.

A detailed theoretical analysis of the electron-nuclear spin
dynamics  still represents a challenging
problem~\cite{KozlovJETP07,ChristPRB07}. It can be discussed
qualitatively taking into account the different stages of the DNP
formation which occur on very different time scales. In the early
stage, the dynamics of the nuclear polarization is determined by
the fast precession of the electron spin about $B_f$ with a period
in the nanoseconds range~\cite{MerkulovPRB02}. Each polarization
event interrupts this precession, resulting in transfer of angular
momentum between the electron and nuclear spin systems. The
non-uniformity of the electron density distribution does not play
an important role at this stage because it gives rise to much
slower processes. E.g., it causes a slow precession of the NSF and
consequently electron spin relaxation on a characteristic time
scale three orders of magnitude longer than that of the electron
spin relaxation in the ``frozen'' NSF~\cite{MerkulovPRB02}.

As we address in effect the early DNP stages, we may use in our
analysis of the experimental data a simplified approach based on
the ``box'' model as well as its
generalization~\cite{Semenov83,KozlovJETP07,KozlovX}. The goal of
our analysis is to compare qualitatively the dynamics of the
nuclear spin polarization and that of the suppression of the
transverse NSF components. The simple version of the box model
assumes the electron density to be uniformly distributed over the
QD volume~\cite{Semenov83,KozlovJETP07,KozlovX}. It gives
$B_N$-fields whose strengths are comparable to the fluctuating
nuclear field $B_f$ only. Our experimental results agree well with
this result: $B_f = 30$~mT (at weak excitation keeping the nuclear
system non-polarized) and $B_N = 40$~mT, see Fig.~7(a). The simple
box model also predicts that, in spite of the small achievable
nuclear spin polarization, the transverse component of $B_f$ may
be considerably decreased by optical pumping, also in agreement
with our results.

To understand the underlying physics, we have performed model
calculations using the simple and extended box models. Details of
these models are given in Appendix B. The effective magnetic
fields of the DNP and NSF destroying the electron spin
polarization are given by the
equations~\cite{Zakharchenya,PetrovPRB08}:
\begin{eqnarray}
B_N &=& \frac{A_{hf}}{g_e \mu_B}\left<\hat I_z\right>,\nonumber\\
B_f &=& \frac{A_{hf}}{g_e \mu_B}\sqrt{2\left<\hat I^2_x+\hat I^2_y\right>}.
\label{eqn26m}
\end{eqnarray}
Here $A_{hf}$ is the constant of the hyperfine electron-nuclear
spin interaction and $\left<\hat I_{\alpha}\right>$ ($\alpha =x$,
$y$, $z$) is the averaged value of the $\alpha$-component of the
total nuclear spin.

When the nuclei are non-polarized, $\left<\hat I_z\right> = 0$ and
$\left<\hat I^2_x+\hat I^2_y\right> = 2/3\left<\hat I^2\right>$.
Assuming $I_i = 1/2$ (in units of $\hbar$) for all nuclear spins,
$\left<\hat I^2\right> = (3/2) N$ where $2N$ is the number of
nuclei interacting with the electron. From these relations follows
$\left(B_f\right)_{non}=A_{hf}/\left(g_e\mu_B\right)\sqrt{2N}$ for
the non-polarized nuclei.

From the box model calculations we can derive the strongly
polarized state of the nuclear spin system achievable under strong
(but short) optical pumping. The average value of $\hat I_z$ for
this state is given by:
\begin{equation}
\left<\hat I_z \right> \approx 2\sqrt{\frac{N}{\pi}}.
\label{eqn27m}
\end{equation}
The average polarization of the nuclear spin system remains small.
Therefore the effective magnetic field of the DNP is also small:
$B_N \approx A_{bf}/(g_e \mu_B) 2\sqrt{N/\pi}$, which is about the
effective field of the NSF for a non-polarized nuclear system. At
the same time, the transverse component of the NSF tends to be
considerably reduced in the strongly polarized state:
\begin{equation}
\left<\hat I_x^2\right>+\left<\hat I_y^2\right>= \left<\hat I_z\right>.
\label{eqn28m}
\end{equation}
From this result it follows that the transverse component of the
NSF is reduced to: $ \left(B_f\right)_{pol}= A/(g_e
\mu_B)2\sqrt[4]{N/\pi}$, which is small compared to the initial
value $\left(B_f\right)_{non}$. For instance,
$\left(B_f\right)_{pol}/\left(B_f\right)_{non} \approx 0.07$ for
$2N=10^5$ nuclei.

The assumption about homogeneity of the electron density used in
the box model is not valid in real QDs. Due to the inhomogeneity
of the density, nuclear spin sub-systems with different $I$
interact with each other (via hyperfine interaction with the
electron) which results in a larger achievable spin polarization
than the one predicted by the box model. In
Ref.~\onlinecite{KozlovX}, an extended model in which the electron
density is approximated by a series of step functions (``graded
box'' model) is considered. The upper limit of the nuclear spin
polarization predicted by this model is: $\left<\hat I_z \right>
\propto \sqrt{kN}$ where $k$ is the number of steps in the
function. The graded box model also predicts that
$\left(B_f\right)_{pol}/\left(B_f\right)_{non} \propto
\sqrt[4]{k/N}$. A limited number of steps $k << N$ should give an
appropriate description of the electron wavefunction/density
distribution in the studied annealed QDs. Still this extended
model predicts a nuclear spin polarization, which is only several
times larger than the one predicted by the simple box model.
Similar to the simple box model, it also predicts a strong
suppression of the NSF. These results support the main conclusions
derived from the simple box model. Another verification of the box
model has been done by Zhang et al.~\cite{ZhangPRB06} using
numerical simulations for a limited number of nuclei.

From these conclusions, we may consider the following scenario of
the nuclear spin dynamics. The optical pumping rapidly polarizes
the nuclei within sub-systems each with a fixed $I$, thus creating
relatively small DNP and, at the same time, strongly suppressing
the transverse components of the NSF. Further pumping gives rise
to a slow increase of the DNP. In our experiments, however, this
long-term buildup is prevented by the partially modulated
polarization of the pumping. We also note that the strong
suppression of the NSF effect favors formation of an
electron-nuclear spin polaron as reported in
Ref.~\onlinecite{OultonPRL07}.

Dynamic nuclear polarization is usually treated in terms of
lowering the nuclear spin temperature~\cite{Zakharchenya}. The
small effective magnetic field $B_N$ which we observed for
$\sigma^+$- and $\sigma^-$-polarized excitation means that the
decrease of the nuclear spin temperature is small for our
experimental conditions. At the same time, the considerable
suppression of the NSF effect for the same conditions indicates a
strong modification of the nuclear spin dynamics. We emphasize
that an equally efficient suppression of the NSF effect is
observed for the $\sigma^+/\sigma^-$ excitation protocol [see
Fig.~7(b)] when no spin temperature decrease is expected.
Therefore we conclude that the concept of an effective temperature
cannot adequately describe the spin state of the nuclear system in
our case and we should describe it in the frame of a decrease of
the entropy of the system. We believe that this problem deserves
further theoretical investigations.

\section{conclusion}

The experiments performed here have allowed us to determine the
effective magnetic field of the nuclear spin fluctuations in an
ensemble of (In,Ga)As quantum dots. This NSF field is about 30 mT at
weak excitation and considerably decreases with pumping. We explain
this effect by a decrease of the transverse component of the nuclear
spin fluctuations under optical excitation with modulated
polarization helicity.

We also considered the time variations of the nuclear spin
fluctuations to explain the electron spin polarization at weak
excitation. These variations can partly explain the increase of
the electron polarization dip width and depth with decreasing
pumping although further investigations are required for a
quantitative modeling of the experiment.

We compared the NSF field with that of the DNP appearing for
very similar experimental conditions. The effective magnetic
field of the polarization observed under strong excitation for a
rather short time is about 40~mT which is by orders of magnitude
smaller than the theoretically predicted upper equilibrium limit.

Both the strong suppression of the NSF and the small DNP are
satisfactorily explained in the framework of the box model for the
electron-nuclei hyperfine interaction. The optical pumping rapidly
polarizes the nuclei within sub-systems consisting of nuclear
states with fixed total angular momentum. This process results in
a strong suppression of the transverse components of the NSF and
in a relatively small average polarization of the nuclear spins.
Further nuclear polarization is a much slower process with has not
been realized for the experimental conditions in the present
studies.

\section*{ACKNOWLEDGMENTS}

The authors thank I. Ya. Gerlovin for fruitful discussions. The work
was supported by the Deutsche Forschungsgemeinschaft (GK726 and
Grants Nos. BA1549/11-1 and 436 RUS 17/144/05) and the BMBF research
program ``nanoquit''. The work was partly supported by ISTC (grant
2679), by the Russian Ministry of Science and Education (grant
RNP.2.1.1.362) and by the Russian Foundation for Basic Research. R.
Oulton thanks the Alexander von Humboldt foundation for support.

\appendix

\section{Calculation of electron spin polarization for frozen and
melted NSF}

\subsection{Frozen NSF}

Equations~(\ref{eqn4},\ref{eqn5}) allow us to calculate $\left<
S_{zf} (B) \right>$ as function of $\Delta_x$ and $\Delta_z$. We
keep $\Delta_z$ fixed for simplicity and assume that the optical
excitation gives rise to reducing the distribution parameter
$\Delta_x$. To connect its value with the power density of the
optical excitation, we consider an effective number of nuclear
spins, $N_x$, with non-zero $x$-projection: $N_x=I_x N_L$, where
$I_x=\bigl[\left<\hat I_x^2\right>\bigr]^{1/2}$ is the averaged
$x$-projection of a nuclear spin defined in a standard way via the
spin operator $\hat I_x$ squared. $N_x$ determines the
distribution parameter according to: $\Delta_x \propto
\sqrt{N_x}$. The rate equation for $N_x$ has the form:
\begin{equation}
\frac{dN_x}{dt}=-\gamma_{exc} N_x + \gamma_N (N_{x}^0-N_x)
\label{eqn6a}
\end{equation}
Here $\gamma_{exc}$ is the excitation rate which is proportional
to the pump power $P$, and $\gamma_N$ is the relaxation rate of
the nuclear spin polarization. $N_{x}^0$ is the $N_x$ in absence
of excitation. From this equation we derive the power dependence
of $\Delta_x$:
\begin{equation}
\Delta_x=\frac{\Delta_x^0}{\sqrt{1+\gamma_{exc}/\gamma_N}}.
\label{eqn7a}
\end{equation}
$\Delta_x^0$ ($\propto \sqrt{N_{x}^0}$) is the initial value of
$\Delta_x$ which equals to $\Delta_z$ in the frame of our model. We
then exploit that $\gamma_{exc} \propto P$, make the substitution
$\gamma_{exc}/\gamma_N = k P$ in Eq.~(\ref{eqn7a}), and use the
parameter $k$ to fit the calculations to the experiment.

\subsection{Melted NSF}

Let us consider the population dynamics of the spin-split electron
states under excitation by a long circularly polarized pulse:
\begin{eqnarray}
\frac{d n^+}{dt}&=&-\gamma_{exc} n^+ - \frac{\gamma_e}{2}(n^+ - n^-),\nonumber\\
\frac{d n^-}{dt}&=&\gamma_{exc} n^+ + \frac{\gamma_e}{2}(n^+ - n^-).
\label{eqn8a}
\end{eqnarray}
Here $n^+$ and $n^-$ are the populations of the
$\left|+\frac{1}{2}\right\rangle$ and
$\left|-\frac{1}{2}\right\rangle$ spin states of the resident
electron, respectively. The total population is normalized: $n^+ +
n^- = 1$. The excitation rate $\gamma_{exc}=\gamma_{exc}^0 P$,
where $P$ is the power density of excitation and $\gamma_{exc}^0$
is the excitation rate at $P_0=1$~W/cm$^2$.
Equations~(\ref{eqn8a}) are written for the case when a circularly
(say $\sigma^+$) polarized light pulse depopulates the
$\left|+\frac{1}{2}\right\rangle$ state and populates the state
$\left|-\frac{1}{2}\right\rangle$. $\gamma_e$ denotes the spin
relaxation rate due to time variations of the NSF, which tends to
decrease the population difference $\Delta n = n^+ - n^-$.
Solution of Eqs.~(\ref{eqn8a}) leads to the following time
dependence of $\Delta n$ during excitation:
\begin{equation}
\Delta n(t)=\left(\Delta n_0 + \frac{\gamma_{exc}}{\gamma}\right)%
            \exp(-\gamma t)-\frac{\gamma_{exc}}{\gamma},
\label{eqn9a}
\end{equation}
where $\gamma = \gamma_e + \gamma_{exc}$. $\Delta n_0$ is the
initial population difference, which was created by the previous
long pulse with opposite circular polarization and has partially
relaxed during the dark time between the pulses. To determine
$\Delta n_0$, we take into account the duration of each long
pulse, $\tau_p=15$~$\mu$s, and of the dark time,
$\tau_d=3$~$\mu$s, and also the fact that $\Delta n_0$
periodically changes its sign while having the same absolute value
after $\sigma^+$ and $\sigma^-$ polarized pulses. Simple
calculations give the expression:
\begin{equation}
\Delta n_0=\frac{\gamma_{exc}}{\gamma}%
\left(\frac{1-\exp(-\gamma\tau_p)}{\exp(\gamma_e\tau_d)+\exp(-\gamma\tau_p)}\right).
\label{eqn10a}
\end{equation}

We measured the spin polarization during the second half of a long
pulse from $\tau_p/2$ to $\tau_p$. Therefore, we have to integrate
Eq. (\ref{eqn9a}) over this time interval. Finally, using $S_z
\propto \Delta n$, we obtain expression (\ref{eqn11}) for the spin
polarization.

\section{Description of the simple and graded box models}

The box model has been described in detail in
Ref.~\onlinecite{KozlovJETP07}. It assumes that the
electron-nuclear spin system is comprised of an electron spin
$S=1/2$ coupled to an even number $M = 2N$ of identical nuclei,
each with spin $I_j=1/2$ ($j=1 \ldots M$). The homogeneous
distribution of the electron density over the nuclei results in an
identical interaction of each nucleus with the electron. This high
symmetry of the system gives rise to a Hamiltonian which depends
on the {\em total} angular momentum of the nuclear spins:
$\mathbf{\hat{I}}=\sum_{j=1}^{M}\mathbf{\hat{I}}_j$ (in units of
$\hbar$). The Hamiltonian has the form:
\begin{equation}
\hat{H}=A_{hf} \hat S_z \hat I_z + \frac{A_{hf}}{2}(\hat S_+ \hat I_- + \hat
S_- \hat I_+). \label{eqn26b}
\end{equation}
Here the standard notations for the $z-$projections and rising and
lowering operators of the electron and nuclear spins are used. The
parameter $A_{hf}$ is the constant of the hyperfine interaction
which does not depend on the nucleus number.

Due to the simple form of the Hamiltonian, it commutes with $\hat
I^2$. This gives rise to an {\em angular momentum conservation
law} which is a specific property of the box model. The density
matrix of the nuclear spin system consists of blocks corresponding
to states with fixed total nuclear angular momentum $I$. These
blocks are mutually independent, i.e., the hyperfine interaction
of the nuclear spin system with a polarized electron spin may
transfer the system from one state to another within the same
block only, rather than between different
blocks~\cite{KozlovJETP07}. In other words, the total nuclear spin
system is divided into multiple non-interacting sub-systems with
different momenta $I$.

Due to the conservation law, the average value, $\left< \hat I^2
\right>$, is not changed under optical pumping and equals to that
for the totally non-polarized nuclear spin system described by the
high-temperature density matrix. The latter can be easily calculated
using equations derived in Ref.~\onlinecite{KozlovJETP07}:
\begin{equation}
\left< \hat I^2 \right> = \sum_{I=0}^N W_I I(I+1). \label{eqn15b}
\end{equation}
Here $W_I$ is the probability to obtain the eigenvalue $I(I+1)$:
\begin{equation}
W_I = 2^{-2N} (2I+1)\Gamma_N(I). \label{eqn16b}
\end{equation}
The factor $2^{-2N}$ describes the equal probability of all states,
($2I+1)$ is the number of eigenstates in a block with total angular
momentum $I$, and $\Gamma_N(I)$ is the number of these blocks. The
distribution function $\Gamma_N(I)$ has the asymptotic
expression~\cite{ZhangPRB06,KozlovJETP07}:
\begin{equation}
\Gamma_N(I)=-\frac{2^{2N}}{\sqrt{\pi N}}\frac{d}{dI}\exp%
\left(-\frac{I^2}{N}\right) \label{eqn17b}
\end{equation}
Equations~(\ref{eqn15b},\ref{eqn16b},\ref{eqn17b}) give rise to
the simple result:
\begin{equation}
\left< \hat I^2 \right> = \frac{3}{2}N \label{eqn18b}
\end{equation}
which is valid for all polarization states of the nuclear spin
system.

When the nuclei are non-polarized, $\left< \hat I_z \right> = 0$ and
non-zero value of $\left< \hat I^2 \right>$ in Eq.~(\ref{eqn18b}) is
due to the NSF so that $\left< \hat I^2_{\alpha} \right> = N/2$ for
$\alpha = x, y, z$.

Let us consider the strongly polarized state of the nuclear spin
system achievable in the frame of the box model. Strongly
polarized state means that only one state is populated in each
block. Correspondingly, the density matrix has the form:
\begin{equation}
U_{LI} = \begin{cases} 0 &\text{when $L \ne I$},\\
2^{-2N}(2I+1) &\text{when $L = I$}.
\end{cases}
\label{eqn20b}
\end{equation}
Here $L$ is the $z$-projection of the total nuclear angular momentum
$I$. The average value of $\hat I_z$ is given by:
\begin{equation}
\left<\hat I_z \right>=\sum_{I=0}^N\Gamma_N(I)\sum_{L=-I}^I U_{LI}L
\approx 2\sqrt{\frac{N}{\pi}}. \label{eqn21b}
\end{equation}
The approximate value of $\left<\hat I_z \right>$ has been
obtained using the asymptotic expression (\ref{eqn17b}). Further,
for each block with a given $I$, we obtain for the averaged square
of the transverse nuclear moment:
\begin{equation}
\left<\hat I_x^2\right>+\left<\hat I_y^2\right>=\left<\hat I^2\right>-\left<\hat I_z^2\right> = %
I(I+1) - I^2 = I = \left<\hat I_z\right>. \label{eqn21b}
\end{equation}
This relation is also valid for the total nuclear spin and has
been already used in the main text [see Eqn.~(\ref{eqn28m})].

The extended box model consider an electron density distribution
in the QD, which consists of a series of step
functions~\cite{KozlovX}. It is intuitive that such model can
approximate the real density distribution well for an appropriate
choice of the number $k$ of steps and their widths in real space.
Here we consider for simplicity equidistant steps. The
electron-nuclei hyperfine interaction within one step is
characterized by a constant, $A_{hf,i}$, where $i = 1 \ldots k$
and all the nuclei are assumed to have the same spin.

The analysis shows~\cite{KozlovX} that the extended box model
gives a larger achievable nuclear spin polarization, $\left<\hat
I_z \right> \propto \sqrt{kN}$. For analysis of the NSF in the
strongly polarized nuclear spin system, Eq.~(\ref{eqn28m}) can be
used which remains valid in the frame of the extended box model.
This fact has been used in the main text to come to a strong
suppression of the NSF.


\begin{thebibliography}{}
\bibitem{Note-Width} The difference in FWHM for the two QD samples is
related to different localization volume of electrons
\protect{\cite{MerkulovPRB02,PetrovPRB08}} as well as to different
excitation efficiencies of the samples. Dependence of the FWHM on
excitation is discussed in Sect. IV.
\bibitem{Note-field} We should note here that the good correspondence
of the fit curves and the experimental data in Fig.~7 is related to
the special timing protocol of the excitation we used. Under strong
excitation with a fixed helicity of light the experimental curve
shows a slower rise of $A_{NCP}$ with increasing $|B|$ which cannot
be approximated by Eqs.~(\ref{eqn13},\ref{eqn14}). We tentatively
attribute this behavior to the large spread of nuclear magnetic
fields in the QD ensemble which slowly develops with time. Our
discussion of this phenomenon will be presented elsewhere.
\end{thebibliography}
\end{document}